\documentclass[aps,prl,twocolumn,showpacs,groupedaddress]{revtex4}

\usepackage{amsmath, amssymb, graphics, graphicx}
\usepackage{textcomp, subfigure}
\usepackage{color}
\usepackage[final]{pdfpages}
\usepackage{sistyle}

\definecolor{linkcolor}{rgb}{0,0,0.6} 
\usepackage[pdftex,colorlinks=true, pdfstartview=FitV, linkcolor= linkcolor, citecolor= linkcolor, urlcolor= linkcolor, hyperindex=true,hyperfigures=true]{hyperref} 

\begin{document}
\title{\bf   Phase transition oscillations induce{d} by a strongly focused laser beam}
\author{Cl\'emence Devailly$*$}
\author{Caroline Crauste-Thibierge}
\author{Artyom Petrosyan}
\author{Sergio Ciliberto}
\affiliation
{Universit\'e de Lyon, Laboratoire de Physique, \'Ecole Normale Sup\'erieure de Lyon, CNRS UMR5672, 
46, All\'ee d'Italie, 69364 Lyon Cedex 07, France}

\begin{abstract}
{We report here the observation of a  surprising phenomenon consisting in a oscillating phase transition which appears   in a binary mixture, PMMA/3-octanone, when this is enlightened by a strongly focused infrared laser beam. PMMA/3-octanone has a UCST (Upper Critical Solution Temperature) which presents a critical point at temperature $T_c=\SI{306.6}{K}$ and volume fraction $\phi_c=12.8~\%$ [Crauste \textit{et al., ArXiv}{1310.6720}, 2012]. This oscillatory phenomenon  appears because of  thermophoretic and {electrostriction} effects and non-linear diffusion. We analyze these oscillations and we  propose a simple model which includes the minimal ingredients to produce the oscillatory behavior.
}
\end{abstract}
\maketitle
%

Phase transitions in binary mixtures are still a widely studied subject, specifically near the  critical point where several 
 interesting and not completely understood  phenomena may appear, among them we recall  the critical Casimir forces \cite{GambassiN},
  \cite{Dean}, confinement effects \cite{Bramwell}, \cite{Joubaud} and out-of-equilibrium dynamics after a quench.
  The perturbation of  the  binary mixtures by mean of external fields is also an important and recent field of investigation \cite{ReviewPhaseElectric}. For example, a laser can induce interesting  phenomena {in demixing} binary mixtures because  the radiation pressure  can deform the interface between the two phases and it can be used to measure the  interface tension \cite{Delville1}. Depending on the nature of the binary mixtures, laser illumination can also lead to a mixing or demixing transition. In ref.\cite{RadiationPressureMolecularAssembly}, focused infrared laser light heats the medium initially in the homogeneous phase and causes a separation in the LCST (Low Critical Solution Temperature) system. The radiation pressure gradients in a laser beam  also contribute in the aggregation of polymers, thus  producing  a phase transition. {The local heating may induce   thermophoretic  forces  which attract towards the laser beam  one of the binary-mixture components  \cite{ThermoBinary}. }{Other forces like electrostriction can also be involved \cite{Electrostriction}.}

In this letter, we report a new  phenomenon, which consists in an oscillating phase transition induced by a constant illumination from an infrared laser beam in the heterogeneous region of an UCST (Upper Critical Solution Temperature) binary mixture. Oscillation phenomena in phase transition have already been reported in slow cooling UCST \cite{Vollmer2002Manip},\cite{CatesOscillation} but as far as we know, never induced by a stationary    laser illumination. After describing our experimental set-up, we will present the results. Then we will use a very simplified model which contains the main necessary physical ingredients to induce this oscillation phenomenon.

The medium is a binary mixture of Poly-Methyl-Meth-Acrylate (PMMA) (Fluka, analytical standard for GPC) with a molecular weight $M_w=\SI{55900}{g/mol}$ and a polydispersity $M_w/M_n=1.035$ and 3-octanone (sup. 98\%). Both are purchased from Sigma-Aldrich. This binary mixture presents a upper critical solution temperature (UCST)\cite{Xia1996} measured in reference \cite{ArticlePMMA} around {$T_c=\SI{306.6}{K}$} and at the critical PMMA volume fraction  $\phi_c=12.8~\%$. {The phase diagram in volume fraction-temperature presents a high temperature homogeneous region and a low temperature two phases region with a polymer rich and a polymer poor phase.} We prepare solutions at different volume fractions and mix them at 325 K during one night. They are placed in glass cells inserted in a Leica Microscope. Then, the samples are left several hours at room temperature to let the medium have a proper demixion. A laser beam ($\lambda$ = 1064 nm) is focused on each sample \cite{Ruben}. A white light source is also used to illuminate the sample, {whose images are recorded by a fast Mikrotron MC1310 camera\cite{SupplMat}}. 

\begin{figure}[!h]
\begin{center}
\includegraphics[width=0.8\linewidth]{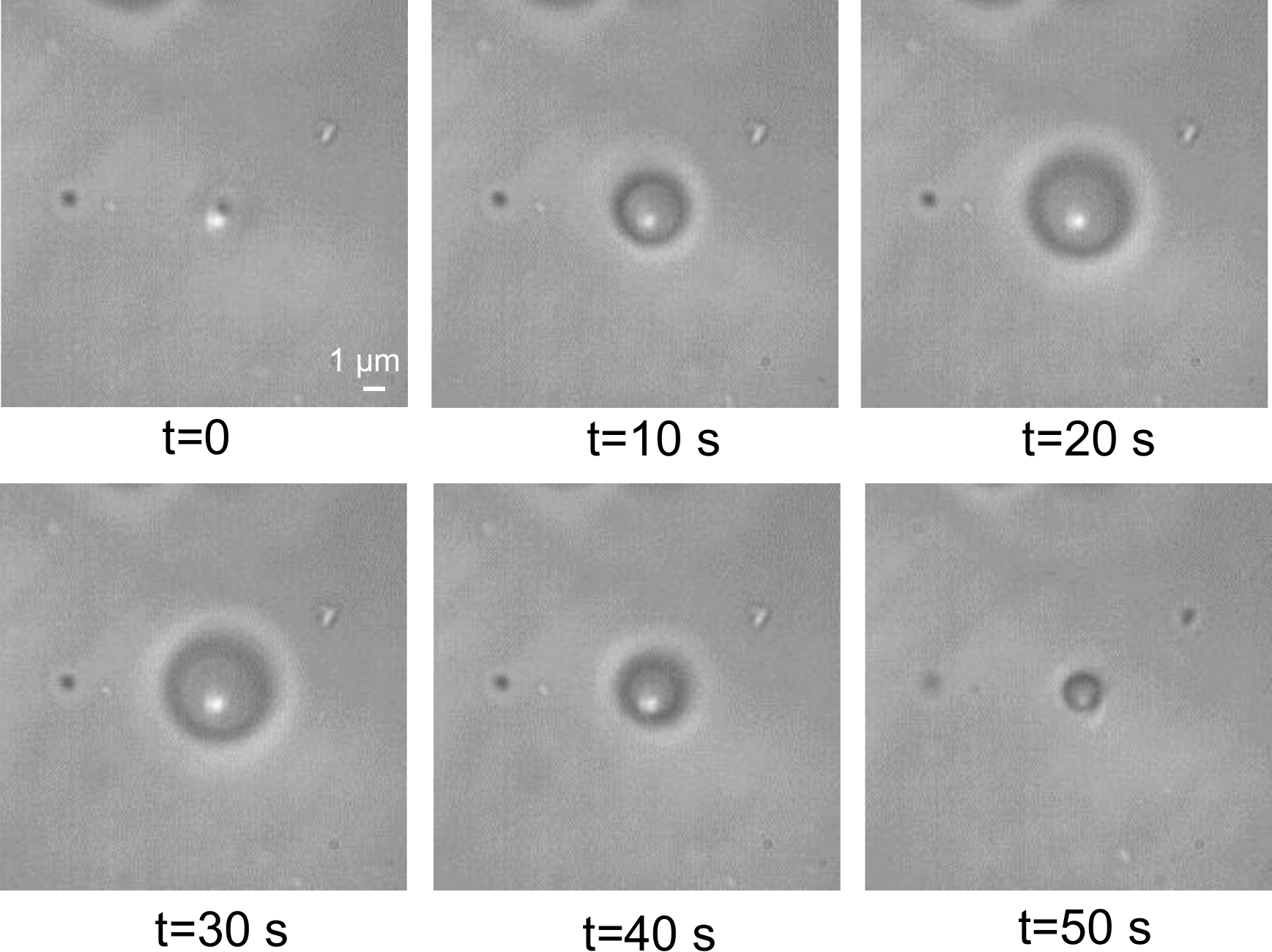}
\caption{{Droplet Oscillation. Images of the octanone/PMMA sample at a volume fraction 12.8 \% in PMMA at room temperature $\SI{298}{K}$. The 6 images have been taken at 10 s time intervals using a microscope objective x63, the image size is $15 ~\mu m$.  A  droplet, rich in PMMA,   growths for the first 30 s and then it decreases.  This oscillatory phenomenon is produced by an infrared laser beam of intensity  \SI{130}{mW}, which is focused inside the sample by the objective.  At the top left of each image, we can see an interface of another PMMA-rich phase which is at equilibrium because at this temperature, the medium is in the heterogeneous phase. The bright point is the reflection of the laser beam.}}
\label{movie}
\end{center}
\end{figure}

At room temperature, the mixture at PMMA volume fraction {${\phi}={\phi_c}=12.8\%$} is in the two phases region. So we can see droplets of one phase in the other one. After a long time, these droplets coalesce. Nevertheless, the medium is thin enough to do not  observe a complete segregation of the two phases due to gravity. At ${\phi}=12.8\%$, theses droplets are steady. When the laser is switched on, a droplet of one phase appears at the focal point of the laser (see  Fig.\ref{movie}). This droplet size increases until a maximum  radius is reached  and then it decreases. When the droplet disappears, another one appears close to  the vanishing one and another cycle of growth and decrease begins. This phenomenon could persist for several oscillations (between 1 and 20). When it stops, we observe some dense { PMMA aggregates on the bottom of the cell}. Sometimes, the phenomenon stops because we place the laser too close to an existing droplet, and the created one coalesces with {an} existing one. 

The oscillatory phenomenon could also appear in a sample with {${\phi}=2\%$} of PMMA, that is to say, a non-critical mixture. Moreover, in this solution, the number of droplets is less important and droplets are more mobile due to a smaller viscosity of the sample. These mobile droplets can easily be trapped by the focused laser beam as particles in an optical tweezers. 
Thus the optical index of the droplets is bigger than the optical index of the bulk. As the two   optical indices are  $n_{PMMA}=1.49$ and $n_{octanone}=1.415$, we conclude that the droplets are the phase rich in PMMA.

{Important  remark: this phenomenon appears when the laser is originally focused in the poor phase at less than $\SI{30}{\mu m}$ of a rich phase. It is probably because one needs a transfer of PMMA materials to create the rich droplet. If the laser is originally on a rich phase, we sometimes observe a growth of the whole rich phase after a transient state where complex phenomena appear (creating interfaces, collapsing).  }

\begin{figure}[!h]
\begin{center}
\includegraphics[width=0.8\linewidth]{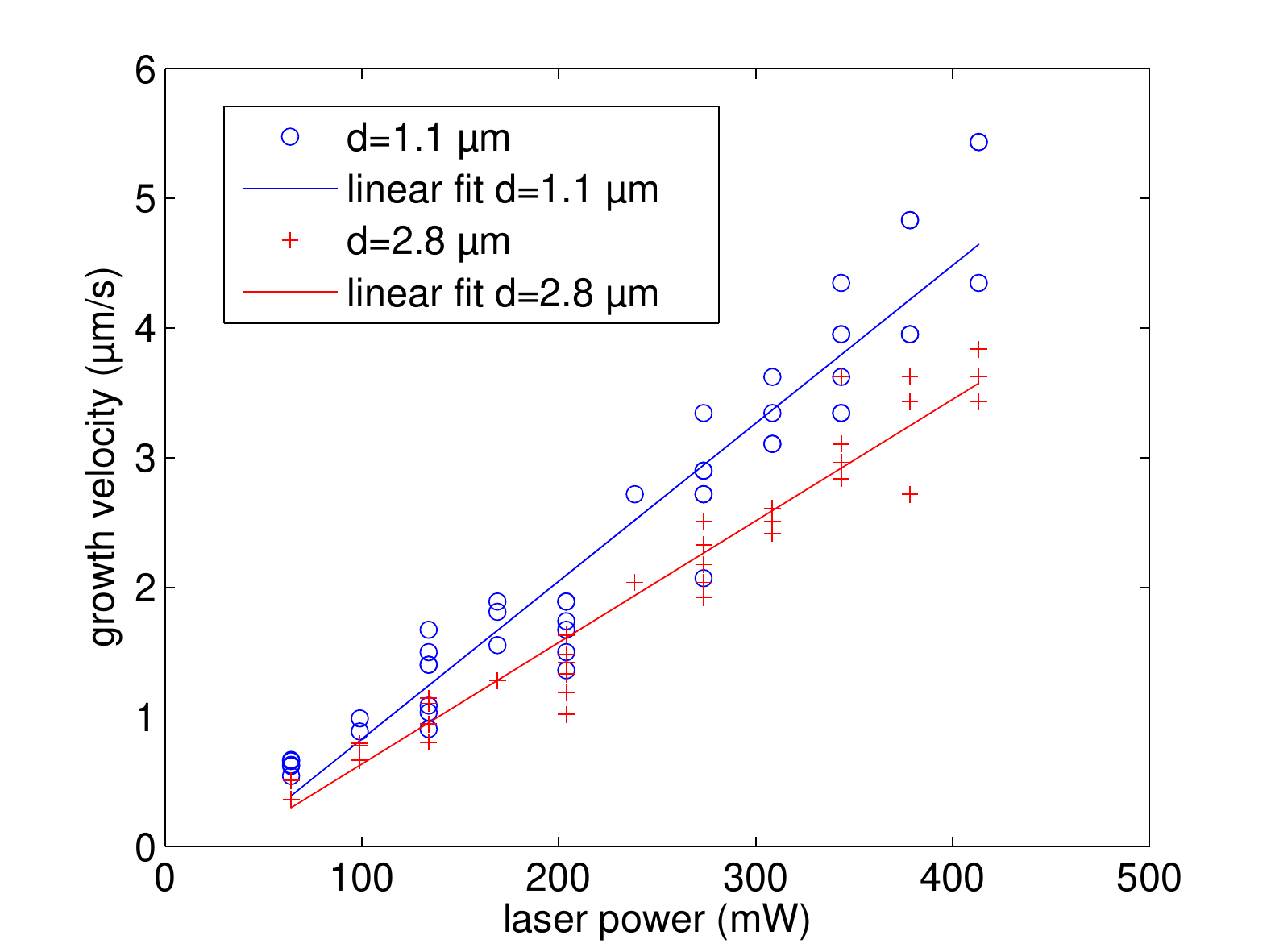}
\caption{Growth velocity of the droplet in the first {instants of their formation as a function} of the laser power.  {
It is measured  from the time which is needed by a growing drop to reach the diameter $d$. The measurement is repeated on various droplets in  different positions of the sample.  The dispersion of the points in the plot  could be due to the heterogeneities of polymer concentration in the cell.} }
\label{growth}
\end{center}
\end{figure}

To characterize this phenomenon, we begin by measuring the growth velocity of droplets as a  function of the laser power in a $\phi=12.8~\%$ sample. The oscillating droplets are acquired at 20 fps with the camera. We measure the time $\Delta t_d$ needed by a growing droplet to reach an imposed diameter $d$. The mean growth velocity is given by $v_g=\frac{d}{\Delta t_d}$. We plot in Fig.\ref{growth} $v_g$ as  function of the laser power for two chosen diameters $d=\SI{1.1}{\mu m}$ and $d=\SI{2.8}{\mu m}$, for several droplets in different positions of the cell. Below \SI{70}{mW}, there are  no droplets. Above \SI{420}{mW}, { the scattering of the laser beam by the sample is too big to do a correct measurement. In the measurement region, the mean velocity $v_g$ is well approximated by a linear function of  the laser intensity, whose slope $p_d$ is a decreasing  function of $d$, specifically in the figure  at  $d=\SI{1.1}{\mu m}$ $p_{d}~=~\SI{12.2}{\mu m/s/W}$ and  at 
$d=\SI{2.8}{\mu m}$ $p_d=\SI{9.4} {\mu m/s/W}$. }

\begin{figure}[!h]
\begin{center}
\includegraphics[width=0.49\linewidth]{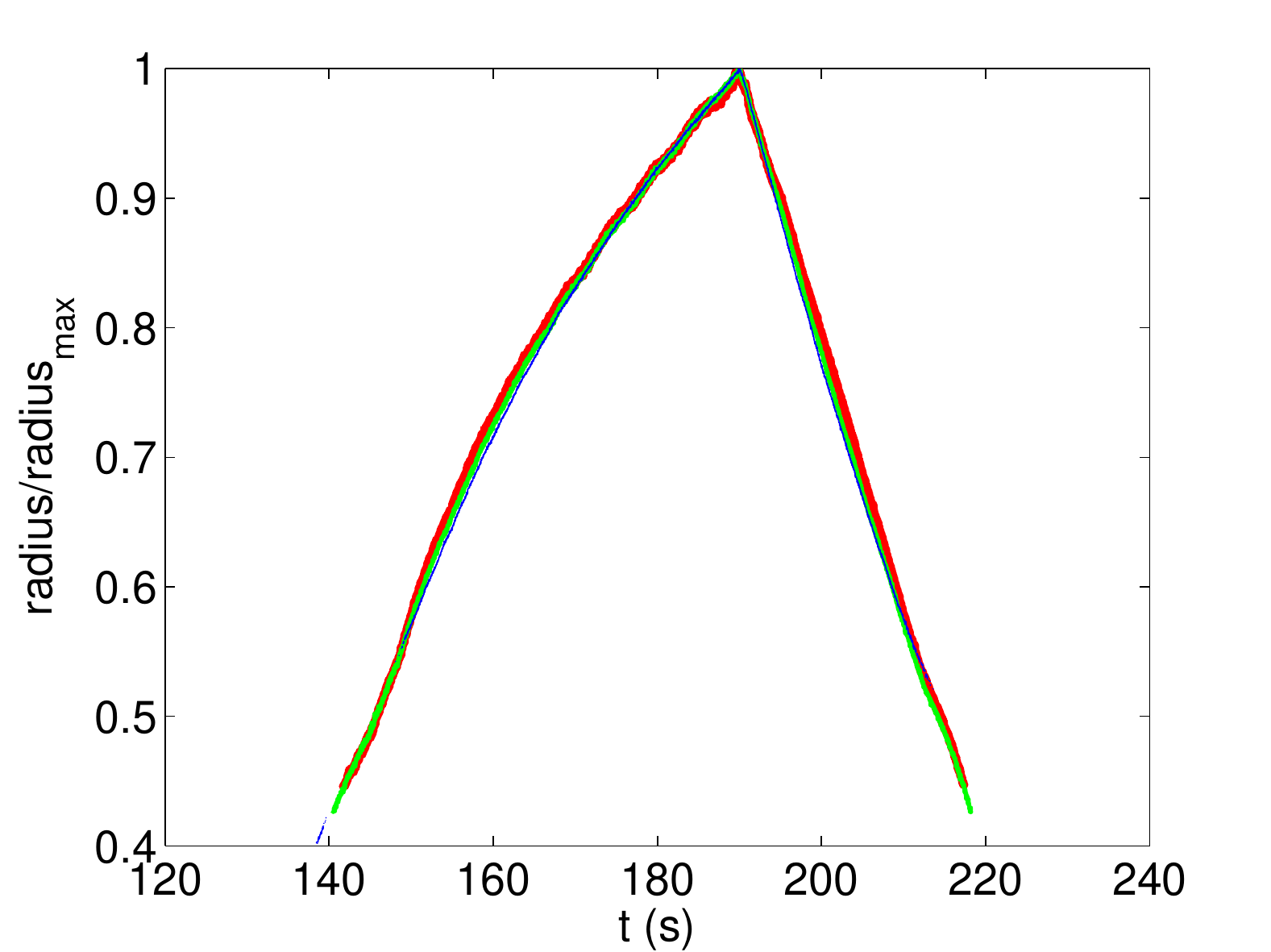}
\includegraphics[width=0.49\linewidth]{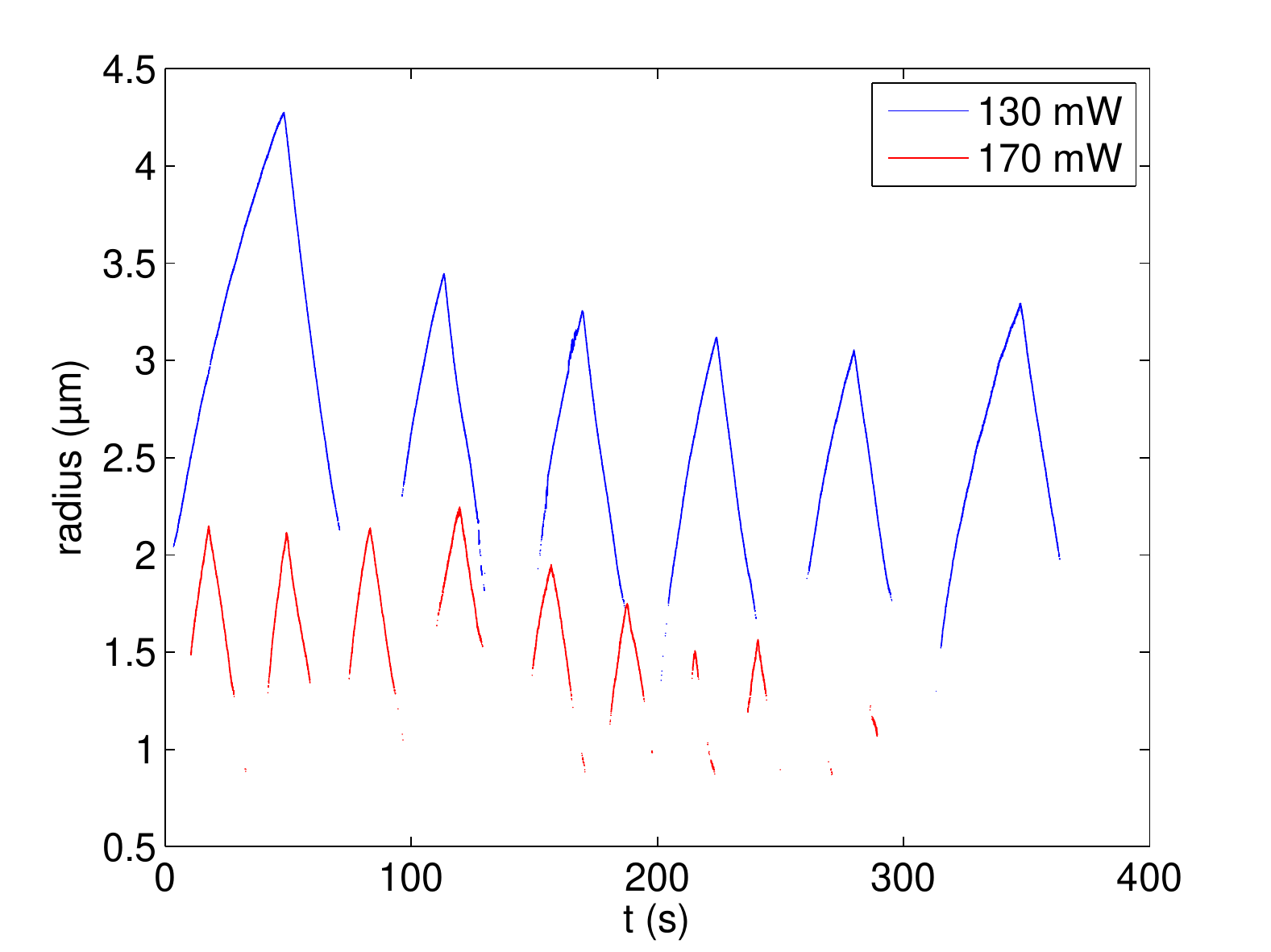}
\caption{(a) Radius of PMMA rich droplets as a  function of time for different detection thresholds (red:155, green:160, blue:168 in grey scale over 255)  of the droplet edge  in the image analysis
normalized to the maximum measured radius. 
The  collapse of all the curves insures that  the time evolution of the radius measured by this method correctly describes  the droplet dynamics.  
The plotted data have been recorded in  a sample with a PMMA volume fraction  $\phi_c=12.8\%$
(laser beam of intensity \SI{134}{mW}). (b) Droplet radius as a function of time  at two different powers (\SI{130}{mW} and  \SI{170}{mW}) of the focused laser beam. The oscillation frequency is a function of the laser power.\label{Oscill}}
\end{center}
\end{figure}

To characterize the oscillations, we implement a program with ImageJ to get the edge of the {droplet} in each image. We then determine the area $\mathcal{A}$ and the radius $R=\sqrt{\mathcal{A}/\pi}$ of the droplet. As the position of this edge is sensitive to the chosen value of the threshold, we plot Fig.\ref{Oscill}-a)  the value of the radius as a function of time for several thresholds. We check that if we rescale the curves by the maximum radius, all curves collapse. The absolute incertitude on $R$ is about \SI{0.5}{\mu m} which is the diffraction limit. {This method is not relevant for too small droplets because we do not get the right edge for two reasons. The first is that diffraction effects dominate. The second is that the contrast inside the droplet  changes during the growth. Thus  a fixed threshold cannot describe the edge of the droplet during all times. Therefore we plot in  Fig.\ref{Oscill}-b)  only the dynamics for the radius above $\SI{1.5}{\mu m}$.}

On one oscillation, we can see that the change of regime (from increase to decrease) is quite sudden. We succeeded in doing at the same spot several oscillations at two different laser intensity. Results are plotted in  figure \ref{Oscill}-b). An increase in laser intensity results in an increase of the  oscillation frequency and a decrease of the maximum amplitude of the droplet. {We tried unsuccessfully to measure quantitatively this effect, because, as we can see in figure \ref{Oscill}-b),  the variation of the droplet maximum radius at fixed intensity perturbs the effect. Furthermore other  parameters, like local changes of concentration, dusts and aging may  disturb our measurement. Nevertheless, from  our analysis at short times, we know that the larger is the  laser power the  faster  is the droplet growth.}

What is the origin of this laser induced transition ? {A droplet of PMMA rich phase is initiated in the PMMA poor phase and then oscillates. Where does this PMMA accumulation come from? } It can not be a simple effect of heating because this binary mixture has a UCST, so an increase in temperature should provoke an homogenization of the solution. But a local heating creates gradients of temperature and then thermophoresis.  {We can estimate the increase of temperature due to the laser by measuring the absorption coefficient of the mixture in a  cell at $\lambda=\SI{1064}{nm}$. To avoid  light scattering, this absorption measurement is performed by keeping the cell at a temperature larger than $T_c$,  when the sample  is well mixed. This gives us an estimation of the extinction coefficient of the mixture $[\epsilon_{PMMA}]\approx \SI{9}{m^{-1}}$ and so an estimation of the  temperature increase, which is about $\Delta T\approx \SI{5}{K}$ at the focal point given by the formula in reference \cite{LaserInduceHeating}. 
This increase should be enough to observe a thermophoretic  effect. As  the  sign of the thermophoretic coefficient {(Soret coefficient)} of this mixture is positive \cite{Soret} the  PMMA is attracted to {low} temperature regions.  {So, this does not explain the first growth  of the droplet but it must be an important issue in the oscillations.}

{The presence of the focused laser beam produces a second effect : the trapping of PMMA. We estimate the stability of this trap, taking into account that the radius of gyration of the polymer is about $\Xi_0=\SI{1}{nm}$ \cite{ArticlePMMA}}. At this size, we are in the Rayleigh approximation for light. We calculate the ratio between the scattering force and the gradient force on the particle \cite{FormuleDipolaire} $\mathcal{R}=\frac{F_{scatt}}{F_{grad}}$. This ratio should be less than one to get a stable trap. We got $\mathcal{R}=\SI{10^{-7}}{}$. The trap is thus stable. But to trap correctly, the trapping force also needs to be bigger than the thermal forces acting on the PMMA bead. To check that, we have to estimate the Boltzmann factor $\exp (-U{_{grad}}/k_{\mathrm{B}T})<<1$ where $U_{grad}$ is the potential of the gradient force {ref. \cite{Ashkin86}. { In our case, inserting the experimental values in the equation for  $U{_{grad}}$ of ref.\cite{Ashkin86} we obtain $U{_{grad}}\approx \SI{5e{-26}}{J}$, which is much smaller than  $k_{\mathrm{B}}T \approx \SI{4e{-21}}{J}$.}} So even if the trap is stable, the gradient force is not sufficient to trap the polymer.}

{The laser can finally induce electrostrictive forces through its electric field gradient \cite{Electrostriction}. As $n_{PMMA}>n_{octanone}$, this force results in an attraction of the polymer close to the focused laser beam through the osmotic compression of the solute. }

{{To summarize}, the rich phase created by the laser is due to an excess of polymer brought by the laser probably by electrostrictive forces.  {As at this excess concentration, the  polymer  mixture  is not thermodynamically stable, the mixture separates  spontaneously and the new phase growths gradually with an increasing amount of PMMA. But this effect is balanced by thermophoretic effect which brings the polymer toward the low temperature region. However this simple explanation does not give us the reason of why the phase should decrease at a certain point. That needs a more precise model, which includes as activation  mechanisms a combination  of the above mentioned thermophoretic effects and electrostriction and their dependences in concentration of PMMA}. }

The oscillation phenomenon in phase transition was already observed in reference \cite{Vollmer2002Manip,CatesOscillation}  in an UCST transition during a slow cooling of the mixture. A simple but powerful model was developed in ref.\cite{CatesOscillation,UltraSlowCooling} based on the Landau phase transition theory. They were able to explain their oscillatory behavior  with the following equation :
\begin{equation}
\partial_t \varphi(x,t)=\partial_x \left[(3\varphi^2-1)\partial_x \varphi\right]
- M^2 \partial_x^4 \varphi-\xi \varphi.
\label{EqVollmer}
\end{equation}
{where $\varphi =(\phi - \phi_c)/(\phi_0- \phi_c)$. $\phi_0(T)$ is the equilibrium {volume} fraction of PMMA as a function of temperature. So for each temperature, $\varphi=1$ in the equilibrium rich phase and $\varphi=-1$ in the equilibrium poor phase. In this equation, 
there is the non linear diffusive term $\partial_x \left[(3\varphi^2-1)\partial_x \varphi\right]$, the interface term $- M^2 \partial_x^4 \varphi$ and a source term proportional to $\varphi$ via a pumping coefficient $\xi$ that in the original model was  $\xi \propto \frac{\partial_t \phi_0}{\phi_0}$.} 
 
\begin{figure}[htb]
\begin{center}
\includegraphics[width=0.9\linewidth]{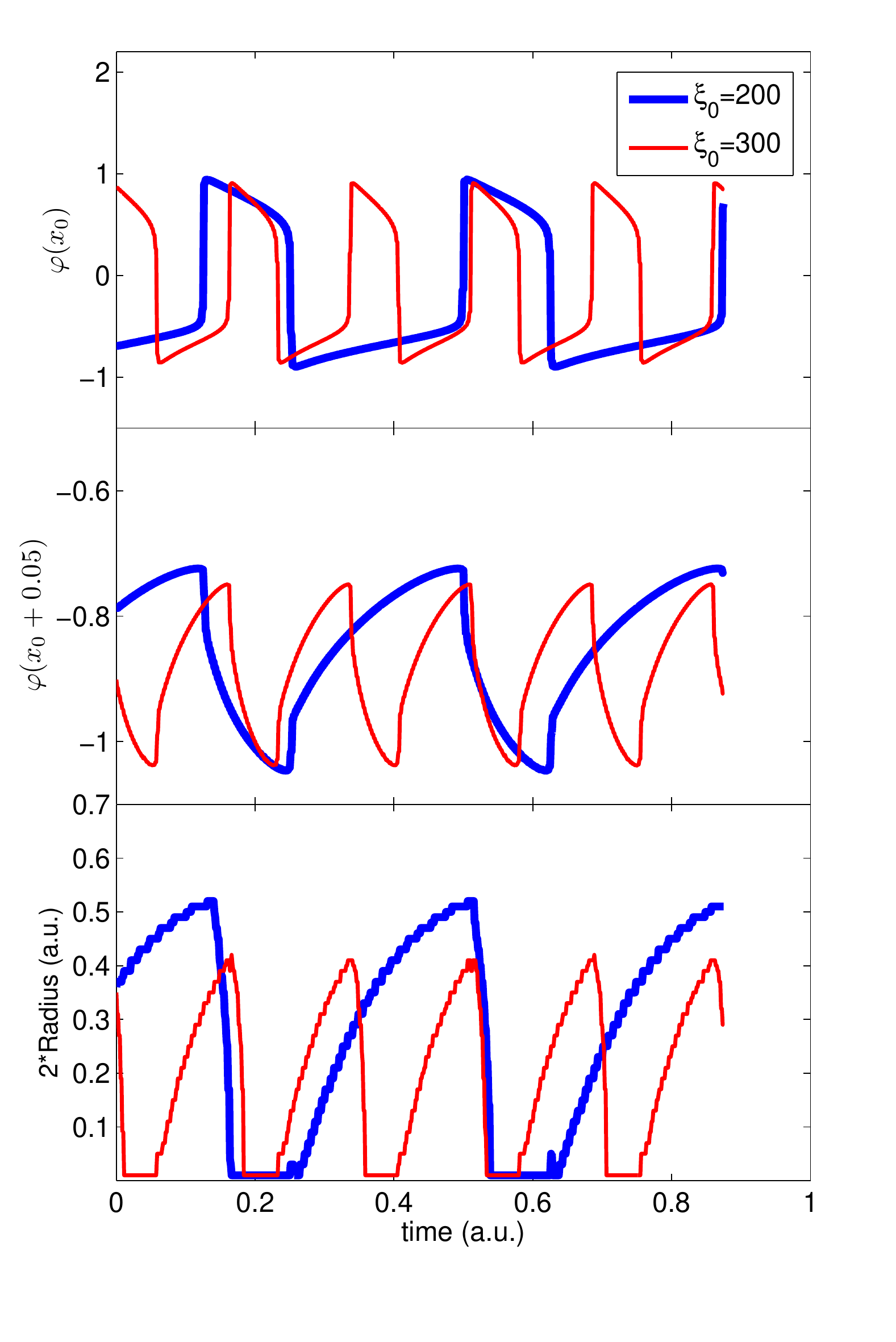}
\caption{{Numerical solution of eq.\ref{EqVollmer}.  Mass-fraction $\varphi$ as a function of time, measured at the forcing point (a) and at $x_o+0.05$  at two different values of $\xi$ (b).  Regular oscillations appear for $\xi$ larger than  a threshold value $\xi_o\simeq 1.5/S_\xi $. The bigger $\xi$,  the faster the oscillations are. (c) Size of the PMMA rich phase calculated numerically from equation \ref{EqVollmer}.   The size is estimated as the  radius 
of the domain in which  $\varphi$ is larger than a defined threshold value, which is $-0.9$ in this figure. {The radius oscillations are the smaller and  the faster  when the source term is bigger}.}}
\label{OscillationNumeriqueConcentration}
\end{center}
\end{figure}

{In order to simulate  the  laser induced oscillations, described in the previous sections, }
we solved numerically eq.\ref{EqVollmer}  with a time independent local source of {amplitude  $\xi=\xi_0$ in a region of size $S_\xi$ around  $x_0$ and $0$ elsewhere. 
 The numerical simulation has been performed \cite{SupplMat} with initial conditions $\varphi=-1$ everywhere, with $M=0.002$ and $S_\xi=0.01$ in the domain $ 0<x<1$ with $x_o=1/2$.   In these conditions, we observe  that for $\xi_o>1.5/S_\xi$  the local concentration oscillates in a region around  the forcing point.  These oscillations can be seen in Fig.\ref{OscillationNumeriqueConcentration}-a)-b) where we plot the value of the local concentration at $x_o$ and at $x_o+0.05$  for two different values of $\xi_o$.}

Looking at  Figs.\ref{OscillationNumeriqueConcentration}-a)-b) we see that { the bigger $\xi_o$ is  the  quicker and smaller oscillations are},  which is in good agreement with experimental results (figure \ref{Oscill}-b). {We also observe an oscillatory creation of a rich phase in the poor one at the forcing point  \ref{OscillationNumeriqueConcentration}-a-b).
Following the experimental procedure we can also estimate  in the numerical simulation  the radius 
of the domain in which  $\varphi$ is larger than a defined threshold value,  fixed for example at $-0.9$. 
The results of this estimation  are  plotted in Fig.\ref{OscillationNumeriqueConcentration}-c) where we see that  the numerical model produces a   time evolution of the radius, which is very similar to that of the experiment.
}
{These results show that with this simplified approach}, we  get the most important ingredients to produce an  oscillatory phenomenon, that is to say : (1) a non linear diffusion term; (2) a source term whose value changes sign when one changes phase.

{ As we have already discussed, the  physical  origin of  this  source term {comes from} 
thermophoresis and electrostriction because both effects contribute to  antagonist changes of the local PMMA density.}  The determination of the relative importance of these two contributions  will require other experimental  setups to precisely measure  the Soret and electrostriction  coefficients.  
Furthermore the source term in  eq.\ref{EqVollmer} is oversimplified with respect to a full model based on the Landau theory in which we consider a local dependence on temperature and electric field of the coefficients. The derivation of this  rather  complex model will be {the subject of  another  report}. The one presented here has the advantage  of being simple and of describing the main effects.  

{{As a  conclusion}, we have  presented  a local oscillating phase transition induced by a focused laser beam. To the best of our knowledge, this phenomenon   has   never been  observed before. We show that a simple  model, based on the Landau theory for phase transition and a local forcing contains enough ingredients to provoke a non-linear oscillatory  behavior. We propose  physical mechanisms which may provoke  the oscillation cycle. The  thermophoresis plays a very important role because the Soret coefficient is positive and PMMA is attracted towards {low} temperature region. In any case it cannot be excluded that other mechanisms (such as electrostricton) may also play a certain role. The connection of our simple and minimal model with a more physical one will be the subject of a theoretical paper where the weight of the various mechanisms will be  evaluated more precisely. 
To improve this study one could  also choose  another model for the phase transition using Flory-Huggins theory \cite{ScalingConcepts}. It is the solution chosen by Anders in  \cite{ThermoBinary}, which could be closest to the experiments because the model is very well suited to describe   polymer-solvent interaction.}

\textit{Acknowledgments}
This paper has been supported by the ERC project OUTEFLUCOP. {We aknowledge fruitful discussions with Jean-Pierre Delville.} \\

$*$ {Present address: School of Physics and Astronomy, University of Edinburgh,
Edinburgh, EH9 3FD, United Kingdom}

\vskip 20pt

\appendix{ \bf Supplementary material for "Phase transition oscillations induce{d} by a strongly focused laser beam''}

\section{\bf Detailled sample preparation and \\  experimental set-up}
We prepare the sample,  under a laminar flow hood, at defined volume fractions by weighting the polymer before adding a volume of 3-octanone calculated from the density of the polymer $\rho_{PMMA}=1.17$ given by the supplier. The solution is then mixed at 325 K during one night to ensure a good dissolution.

The cell containing the sample  is composed by a 1 mm thick glass plate and by a cover slip separated by a $\SI{100}{\mu m}$ thick polycarbonate sheet and glued with NOA 81 under UV light.  The top glass plate has two apertures connected with two filling metallic tubes.  To fill the cell, we heat  all the materials (syringe, needle, cell) and the medium to avoid demixing during the filling. Then, we close the two openings with a small amount of wax. Two metallic tubes avoid { wax to be directly in contact with the mixture}. We leave the cell several hours at room temperature to let the medium demixing properly. {We obtained a cell contaning two phases with typical size of the regions being around $\SI{20}{\mu m}$ to $\SI{200}{\mu m}$.}
{
The filled cell is inserted in a microscope. We use an immersion  oil-objective (Leica x63, N.A. 1.4) to focus the laser beam into 
 the sample. The laser is a  Quantum Forte 700 mW at wavelength 1064 nm supplied by a power source LD3000 to control the power intensity of the beam.  The sample is also illuminated by a white light source.  This light is collected by  the x63 objective and the sample is observed with  a fast camera Mikrotron MC1310. The whole optics of the set-up is given in reference ?? of the article.  The laser power is calibrated by measuring the power of the laser beam just before the objective. Thus, it is not exactly the value of the intensity in the cell. The attenuation of the microscope objective is about 70\% at \SI{1064}{nm}.}

\section{\bf Integration of the model equation}
The numerical integration is performed by finite difference in space, by dividing the interval $0\le x \le 1$ in $N$ points. The integration in time is performed by a fourth order Runge-Kutta method. We checked that the results are independent of $N$ by changing it from $50$ to $400$. The initial conditions is $\varphi=-1$ in all of the points of the interval. Two types of  boundary conditions have been used: a) $\varphi(0)=\varphi(1)=-1$; b) $\partial_x\varphi |_{x=0}=\partial_x\varphi |_{x=1}=0$. We mostly used the a) type because they correspond to have a good reservoir at the extremes, but the results do not change too much using the type b).  The forcing term is 
$\xi_o= \xi_o'/S_\xi$ for $(x_o-S_\xi /2)\le x \le (x_o+S_\xi /2)$ and 0 elsewhere. Notice that there are three length scales in this problem. The integration domain which is set to 1, the forcing size and $M$. The oscillation do no appear if $M>S_\xi$.
The results presented in Fig. 4 have been obtained with $N=100$ and boundary conditions a). More details on the numerical integration of eq. 1 of the article and on its physical background will be the object of a theoretical paper.



\begin{thebibliography}{0}
\expandafter\ifx\csname natexlab\endcsname\relax\def\natexlab#1{#1}\fi
\expandafter\ifx\csname bibnamefont\endcsname\relax
  \def\bibnamefont#1{#1}\fi
\expandafter\ifx\csname bibfnamefont\endcsname\relax
  \def\bibfnamefont#1{#1}\fi
\expandafter\ifx\csname citenamefont\endcsname\relax
  \def\citenamefont#1{#1}\fi
\expandafter\ifx\csname url\endcsname\relax
  \def\url#1{\texttt{#1}}\fi
\expandafter\ifx\csname urlprefix\endcsname\relax\def\urlprefix{URL }\fi
\providecommand{\bibinfo}[2]{#2}
\providecommand{\eprint}[2][]{\url{#2}}

\end{thebibliography}


\begin{thebibliography}{}


\bibitem{ArticlePMMA}
C. Crauste, C. Devailly, A.Steinberger, S. Ciliberto,
{\it arXiv}, 1310.6720 (2013). 

\bibitem{GambassiN}
C. Hertlein, L. Helden, A. Gambassi, S. Dietrich, C. Bechinger,
{\it Nature}, {\bf 451}, 7175 (2008).

%

\bibitem{Dean}
DS. Dean, and A. Gopinathan, 
{\it Jour. Stat. Mech}, {\bf 2009},  L08001 (2009).

\bibitem{Bramwell}
ST. Bramwell, PCW. Holdsworth, JF. Pinton,
{\it Nature}, {\bf 396}, 6711 (1998).

\bibitem{Joubaud}
S. Joubaud, A. Petrosyan, S.Ciliberto, NB. Garnier,
{\it Phys. Rev. Lett.}, {\bf 100}, 18 (2008).

\bibitem{ReviewPhaseElectric}
Y. Tsori,
{\it Rev. Mod. Phys.}, {\bf 81} (2009). 

\bibitem{Delville1}
R. Wunenburger, A. Casner, JP. Delville,
{\it Phys. Rev. E}, {\bf 73}, 036314 (2006)

\bibitem{RadiationPressureMolecularAssembly}
J. Hofkens, J. Hotta, K.  Sasaki, H.  Masuhara, K. Iwai
{\it Langmuir}, {\bf 13}, 3 (1997)

\bibitem{ThermoBinary} 
D. Anders, K. Weinberg
{\it Mech. Mat.}, {\bf 47} 33 (2012).

\bibitem{Electrostriction} 
J.P. Delville, C.Lalaude S. Buil, and A. Ducasse
{\it PRE}, {\bf 59} 5 (1999).


\bibitem{Vollmer2002Manip}
D. Vollmer, J. Vollmer and A.J. Wagner,
{\it Phys. Chem. Chem. Phys.}, {\bf 4 } 1380 (2002).


\bibitem{CatesOscillation} M. E. Cates, J. Vollmer, A. Wagner, D. Vollmer,
{\it Philosophical Transaction}, {\bf 361 } 1805 (2003). 


\bibitem{Xia1996}
{K.Q. Xia and X.Q. An and W.G. Shen}.
J. Chem. Phys., \textbf{105} (1996) 6018--6025.

\bibitem{Ruben} 
J. R. Gomez-Solano, A. Petrosyan, S. Ciliberto, R. Chetrite, and K. Gawedzki.,
{\it  Phys. Rev. Lett.}, {\bf 103 }040601 (2009). 

\bibitem{SupplMat} 
See the supplemental material of this article at URL ??? for complementary information. 

\bibitem{LaserInduceHeating} 
E.Peterman, F. Gittes, C. Schmidt,
{\it Biophysical Journal}, {\bf 84 } 1308 (2003). 

\bibitem{Soret}  By using  another experimental set-up, 
we have  determined  the sign of the Soret effect  of PMMA/3-octanone mixture because its value is unknown in literature. 

\bibitem{FormuleDipolaire} 
K.~C. Neuman, S.~M. Block, 
{\it Rev. Sc. Inst.}, {\bf 75 } 2787 (2004). 

\bibitem{Ashkin86} 
A. Ashkin and J.M. Dziedzic, J.E. Bjorkholm and S. Chu
{\it Optics Letters}, {\bf 11} 288 (1986).

\bibitem{UltraSlowCooling}
J. Vollmer, G.~K. Auernhammer, D. Vollmer,
{\it Phys. Rev. Lett.}, {\bf 98 } 115701 (2007). 

\bibitem{ScalingConcepts}
P.G. de Gennes,
{\it Scaling Concepts in Polymer Physics} (1979). 

\end{thebibliography}
\end{document}